\documentclass[%
 aip,
 amsmath,amssymb,
 reprint,%
floatfix, % added by NK
dvipdfmx, % added by NK
]{revtex4-1}

\usepackage{graphicx}% Include figure files
\usepackage{dcolumn}% Align table columns on decimal point
\usepackage{bm}% bold math
\usepackage[utf8]{inputenc}
\usepackage[T1]{fontenc}
\usepackage{mathptmx}

\usepackage{color}  % added by NK
\usepackage{subfigure}   % added by NK
\usepackage{comment}   % added by NK

\begin{document}

\preprint{AIP/123-QED}

\title
%[REVIEW OF SCIENTIFIC INSTRUMENTS {\bf xx}, xxxxxx (2020)]
{Apparatus for generation of nanojoule-class water-window high-order harmonics}
% Force line breaks with \\

\author{Kotaro Nishimura}
\email[]{kotaro.nishimura@riken.jp}
 \affiliation{Extreme Photonics Research Group, RIKEN Center for Advanced Photonics, RIKEN, 2-1 Hirosawa, Wako, Saitama 351-0198, Japan.}
 \affiliation{Department of Physics, Tokyo University of Science, 2641 Yamazaki, Noda, Chiba 278-8510, Japan.}

\author{Yuxi Fu}
\email[]{K. N. and Y. F. contributed equally.}
\affiliation{Extreme Photonics Research Group, RIKEN Center for Advanced Photonics, RIKEN, 2-1 Hirosawa, Wako, Saitama 351-0198, Japan.}
 \affiliation{Xi'an Institute of Optics and Precision Mechanics, Chinese Academy of Sciences, Xi'an, Shaanxi 710119, P.R. China.}
%\affiliation{Xi???an Institute of Optics and Precision Mechanics, Chinese Academy of Sciences, Xi'an, Shaanxi 710119, P.R.China}

\author{Akira Suda}
\affiliation{Department of Physics, Tokyo University of Science, 2641 Yamazaki, Noda, Chiba 278-8510, Japan.}

\author{Katsumi Midorikawa}
 \affiliation{Extreme Photonics Research Group, RIKEN Center for Advanced Photonics, RIKEN, 2-1 Hirosawa, Wako, Saitama 351-0198, Japan.}

\author{Eiji J. Takahashi}
\email[Correspondence: ]{ejtak@riken.jp}
\affiliation{Extreme Photonics Research Group, RIKEN Center for Advanced Photonics, RIKEN, 2-1 Hirosawa, Wako, Saitama 351-0198, Japan.}

\date{\today}% It is always \today, today,
             %  but any date may be explicitly specified

\begin{abstract}
In our recent study  [Y. Fu {\it et al}., {\it Commun. Phys.} {\bf 3}, 92 (2020)], we have developed an approach for energy-scaling of high-order harmonic generation in water-window region under  neutral-medium condition.
More specifically, we obtained nanojoule-class water-window soft x-ray harmonic beam under phase match condition. It has been achieved by combining a newly developed terawatt-class mid-infrared femtosecond laser and a loose focusing geometry for high-order harmonic generation.  The generated beam is more than 100 times intense compared to previously reported results.
The experimental setup included two key parts: terawatt mid-infrared femtosecond driving laser [Y. Fu {\it et al}., {\it Sci. Rep.} {\bf 8}, 7692 (2018)] and specially designed gas cell.
Despite the dramatic drop in the optimal gas pressure due to loose focusing geometry, it still reached 1 bar level for helium.
Moreover, faster leaking speed caused by larger pinhole size of the gas cell made the use  of a normal gas cell impossible.
Thus, we have designed a double-structured pulsed-gas cell with a differential pumping system, which enabled providing sufficiently high gas pressure. Moreover, it allowed reducing gas consumption significantly. 
Robust energy-scalable apparatus for high-order harmonic generation developed in in this study will enable the generation of over tens nanojoule water-window attosecond pulses in the nearest future.

\end{abstract}

\maketitle

\section{INTRODUCTION}
High-order harmonic generation (HHG) is a promising approach to realize a lab-scale soft x-ray coherent light source with attosecond time resolution~\cite{PhysRevX.7.041030, Saito2016, Gaumnitz:17}, excellent beam quality~\cite{Zhang:04}, and synchronized infrared (IR) pulses with attosecond time jitter~\cite{doi:10.1063/1.3475689, Chini:09, Mashiko:10}. 
Since its discovery over 30 years ago\cite{McPherson:87,Ferray_1988}, an apparatus of HHG has been improved in terms of increasing photon flux, pulse energy, and photon energy. 
Currently, high-order harmonic (HH) pulse energy reaches $\mathrm{\mu J}$-order per pulse in the extreme ultraviolet (XUV) region\cite{Takahashi:02, Takahashi2013} and the photon energy extends to the water-window  (284-543 eV) soft x-ray region\cite{PhysRevLett.101.253901,Cardin_2018,Pupeikis:20}.

A straightforward way to increase HH pulse energy is to increase pulse energy of the driving laser. 
To achieve an efficient conversion of HHs from the high-energy driving laser pulse with energy scaling ability, loose-focusing method~\cite{Takahashi:03, PhysRevA.68.023808, doi:10.1063/1.1637949, PhysRevA.84.053819, Heyl_2012, doi:10.1063/1.4812266, Heyl_2016, Wang_2018, PhysRevA.98.023426, Kov_cs_2019} is widely adopted. It was originally proposed and developed by Takahashi and co-workers in 2002\cite{Takahashi:02}. 
In  loose-focusing method, the pulse energy of HHs can be efficiently scaled up with that of the driving laser under a phase-match condition. 
Thanks to the loose-focusing method, an intense HHG sources have been developed and found applications in single-shot coherent diffractive imaging\cite{Rupp2017}, ptychography of biological samples\cite{Baksheaaz3025}, and nonlinear optics in XUV region\cite{PhysRevA.98.023426}. 
Loose-focusing method with a meter-scale focusing geometry typically requires a terawatt-scale (TW-scale) high power laser system. That is why it has been widely used in Ti:sapphire lasers.

The wavelength of the driving laser is one of the key parameters that determine the limits of HH photon energy. 
The highest HH photon energy, which is called the cut-off energy, is given by 
$h\nu_{\rm cutoff} = I_{\rm p} + 3.17 U_{\rm p}$ at the single atom level\cite{PhysRevLett.68.3535}, where $I_{\rm p}$ is the ionization potential of the atom and  $U_{\rm p} {[\rm eV]} = 9.337 \times 10^{-14} I [{\rm W/cm^2}]\ (\lambda {\rm[\mu m]})^2$ is the electron quiver energy (ponderomotive energy) in the laser field.  Here $I$ and $\lambda$ are intensity and wavelength  , respectively. 
Laser intensity is limited by the ionization degree in order to fulfil phase-matching in HHG. Thus, using the driving laser with longer wavelengths is the best way to extend HH photon energy limits efficiently.
Optical parametric amplification (OPA) lasers are capable of generating longer wavelengths compared to Ti:sapphire counterparts. Many groups capitalized on this advantage and  generated HHs in the water window region\cite{PhysRevLett.101.253901, Popmintchev10516, Xiong:09, PhysRevLett.105.173901, Takahashi2010, Popmintchev1287, Ishii2014, Cousin:14, Silva2015, Teichmann2016, Stein_2016, Li2017, Johnsoneaar3761, Schmidt:18, Cardin_2018, doi:10.1021/acs.jpclett.8b03420, Li:19, Pupeikis:20, Barreau2020,gebhardt2020bright} and used them for various applications\cite{Pertot264,Attar54,doi:10.1021/acs.jpclett.9b03559}.
At the same time, OPA lasers have several drawbacks.  In particular, they feature  low IR pulse energy  (mJ or sub-mJ range) and low conversion efficiency. The latter is caused by non-optimal scaling of the fundamental wavelength, $\propto \lambda^{-6}$ at the single-atom level\cite{PhysRevLett.98.013901,PhysRevLett.103.073902}.
Moreover,  the phase-matching condition becomes stricter with  HH order increasing at the macroscale level.
From the technical point of view, the optimal gas pressure for phase-matching increases dramatically higher with the fundamental wavelength increase. The latter results in significant increase in gas consumption.
Moreover, high-pressure gas target may seriously deteriorate vacuum level along the propagation path of HHs outside of the gas cell\cite{doi:10.1063/1.5041498}.
Currently, HH pulse energy is mostly limited by pJ or even fJ levels that hinders its potential applications in XUV region.

Recently, we have demonstrated the possibility to generate nJ-class water-window HHs\cite{Fu2020} in  loose-focusing geometries under neutral-medium condition. 
We used TW-class femtosecond laser system as a driving laser.  In particular, we applied dual-chirped optical parametric amplification (DC-OPA) technique.  It features broad wavelength tuning range of 1.2-2.4 ${\rm \mu m}$ and excellent energy scalability up to 100 J \cite{Fu2018}. 
 Next, we combined this laser and loose-focusing method. It allowed maintaining robust and tractable phase-matching and neutral-medium conditions and achieving total HH energy of several nJ in the water-window region.

In this paper, we present a novel apparatus for generation of nJ-class HHs in the water-window region.
In order to supply a sufficiently high-pressure gas with a few cm lengths while keeping good vacuum outside of the harmonic generation point, we designed a double-structured pulsed-gas cell featuring differential pumping capability.
We calculated pressure in double-structured pulsed-gas cell and vacuum chamber as a function of time and backing pressure.
The calculation results are in good agreement with the measured values.
The pressure in the interaction region is easy to control by tuning the backing pressure.
The gas pressure in the interaction region can be increased to over 1.5 bar, while a turbo molecular pump operates normally for He.
The gas consumption of pulsed-gas operation is also reduced to 3.55${\%}$ of that of continuum-flow operation. 
Double-structured pulsed-gas cell allows more than ten-fold increase in  HH intensity compared to continuum-gas cell. For the latter,  the highest gas pressure is limited by the gas leak.
Gas pressure dependence in HHG indicates that the high-pressure gas can satisfy phase-matching condition and increase HH intensity.
As a first application of this source, we demonstrated the absorption spectroscopy analysis of aromatic thin film samples.
Our apparatus was able to obtain two-dimensional fine absorption structures near carbon K-edge with a reasonable energy resolution.

\section{Helium phase-matching condition in the water-window region}
In order to generate harmonic beams from gas efficiently, constructive interference between XUV or soft x-ray photons emitted from each gas atom is required. 
This condition is realized by matching the phase between driving laser field and HHs, which is called "phase-matching condition". 
When driving laser and HHs are colinear, the phase difference is described by the wave number, $\Delta k = k_q - q k_1$, where $k_q$ is the wave number of HHs, $q$ is HH order, and $k_1$ is the wave number of the driving laser field. 
Wavenumber $\Delta k$ is associated with neutral gas dispersion, plasma dispersion, and geometric phase difference in the loose-focusing geometry.
Phase mismatch caused by the driving laser modulation is proportional to HH order. Thus, the phase difference becomes very sensitive when the fundamental wavelength is long enough and HH photon energy is sufficiently high. 
Loose-focusing method for HHG ensures that the phase-matching condition is dominantly satisfied  by balancing  geometric phase and  gas dispersion.
The optimal gas pressure ($P_{\rm opt}$) for the phase-matching condition is given by
\begin{equation}
\label{popt}
 P_{\mathrm{opt}} = \left[ r_e N_{\mathrm{L}} \omega_0^2 \left( \frac{2 \pi \delta n}{r_e N_{\mathrm{L}} \lambda_1^2} + \frac{f_1}{q^2} - \eta  \right) \right]^{-1},
\end{equation}
where $r_e$ is the classical electron radius, $N_{\mathrm{L}}$ is the Loschmidt's constant, $\omega_0$ is the beam waist of the driving laser, $\delta n$ is the refractive index difference per one atm between gas and vacuum for the driving laser, $f_1$ is the real part of the atomic scattering factor of gas for HHs, and $\eta$ is the gas ionization degree. 
The perfect phase-matching ($\Delta k = 0$) can be achieved by tuning the gas pressure in the positive and finite region.  In this case, the ionization degree should satisfy the following inequality:
\begin{equation}
\label{etacrit}
 \eta < \frac{2 \pi \delta n}{ r_e N_{\mathrm{L}} \lambda_1^2} + \frac{f_1} {q^2}.
\end{equation}

When the left-hand side and the right-hand side of  (\ref{etacrit}) are equal , the optimal pressure blows up. The corresponding  ionization degree is called "critical" ionization fraction \cite{PhysRevLett.83.2187}.
When the fundamental wavelength is 1.55 $\rm\mu m$ and the HH photon energy is 284 eV, the critical ionization fraction of He becomes 0.12$\%$, which is smaller than the value of 0.5$\%$ in HHG driven by Ti:sapphire laser.
In loose-focusing method, the focused intensity is set to be moderate values to ensure
the ionization degree is below the critical ionization fraction.
As a result, the phase-matching condition becomes robust and tractable in neutral gas media. In this media, the  ionization degree is relatively low and homogeneous in time and space.

\begin{figure}[htbp]
\includegraphics[width=0.73\linewidth]{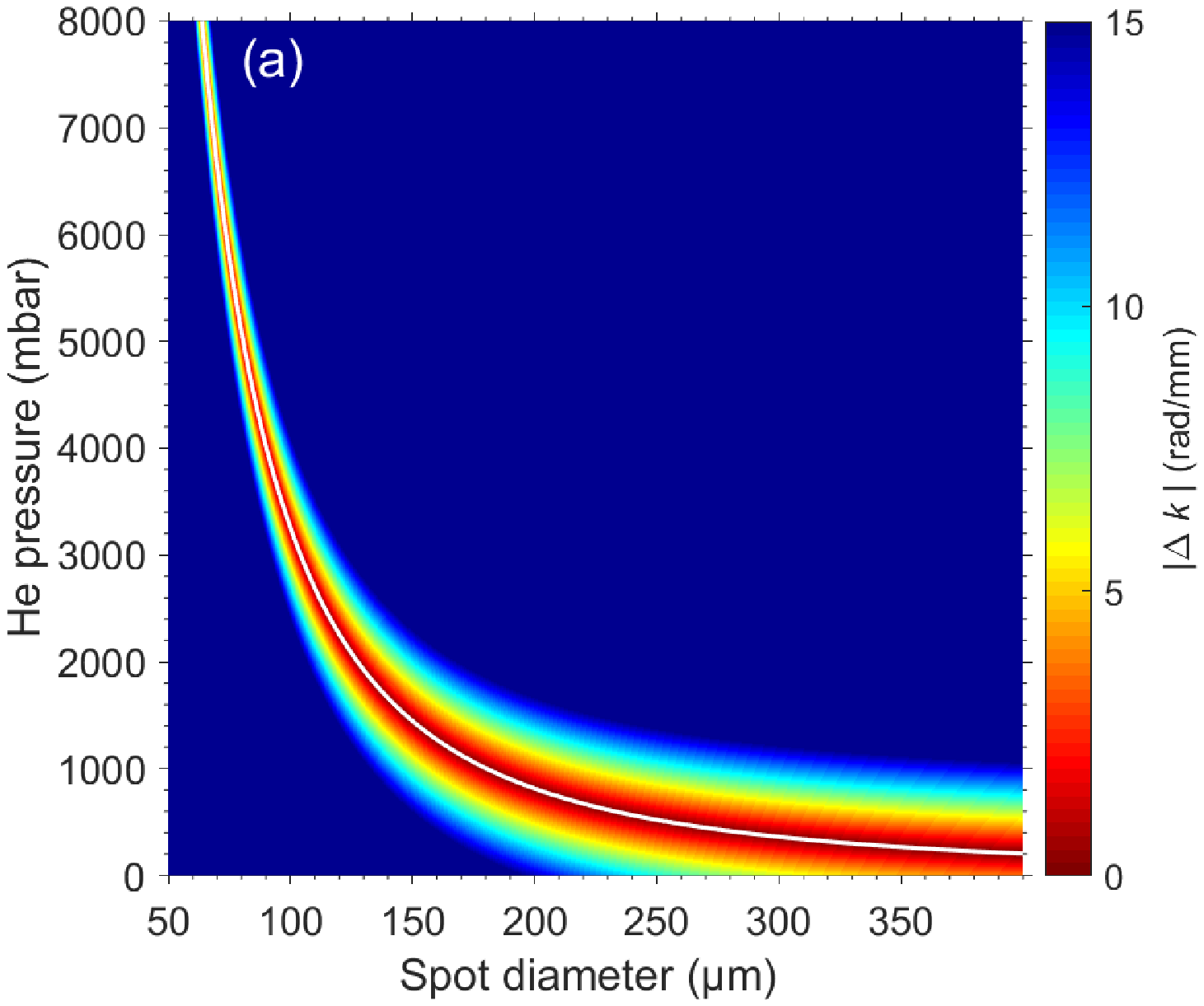}
\includegraphics[width=0.73\linewidth]{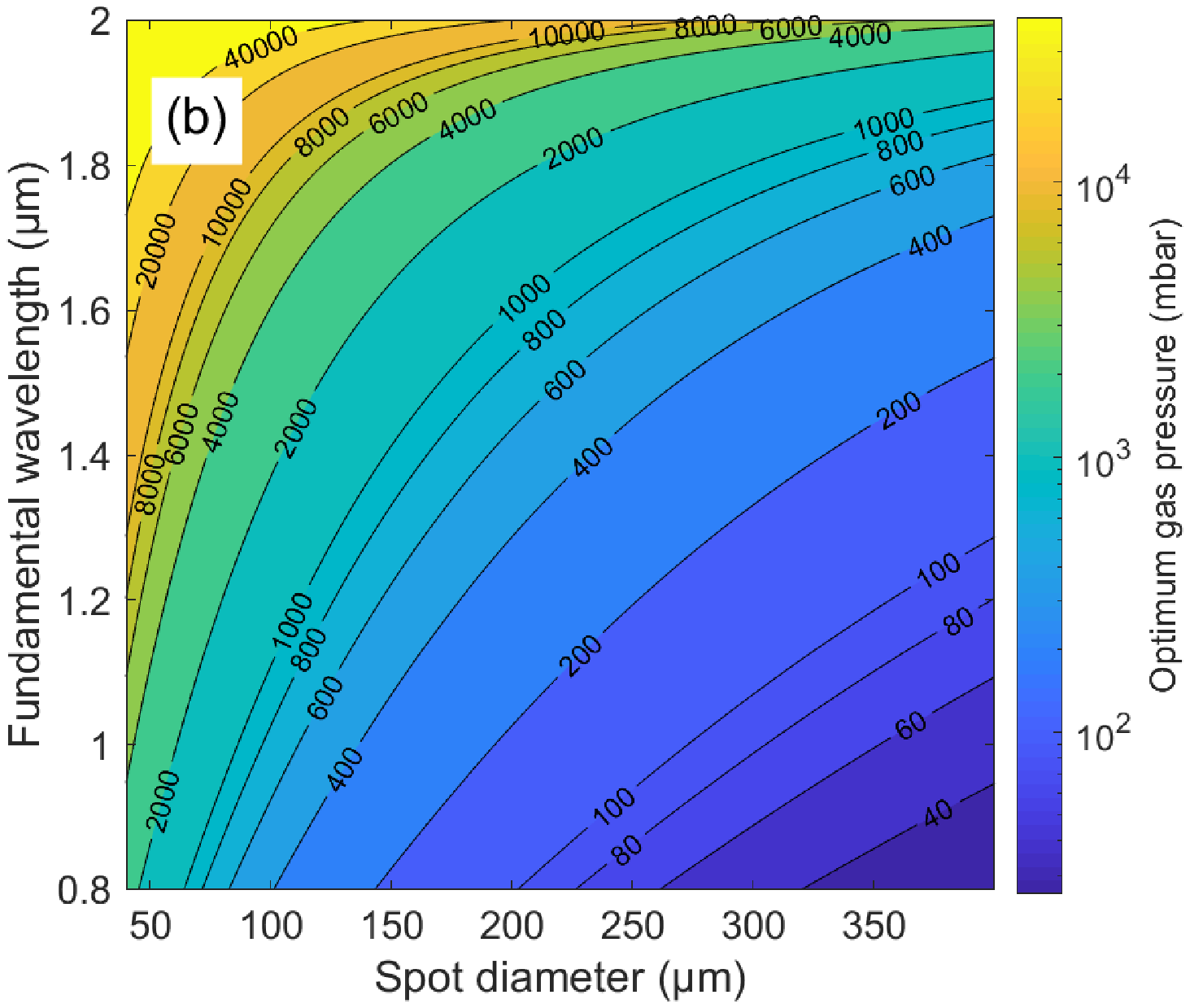}
\caption{\label{fig:popt} (a) Phase mismatch as a function of spot diameter and He pressure.  The fundamental wavelength is 1.55 $\rm\mu m$, the HH photon energy is 284 eV, and the ionization degree is 0.07$\%$. The white curve corresponds to the optimal pressure (zero phase mismatch).
 (b) Optimal pressure as a function of spot diameter and fundamental wavelength in the cut-off region.
 }
\end{figure}

\begin{figure*}[htb]
\includegraphics[width=\linewidth]{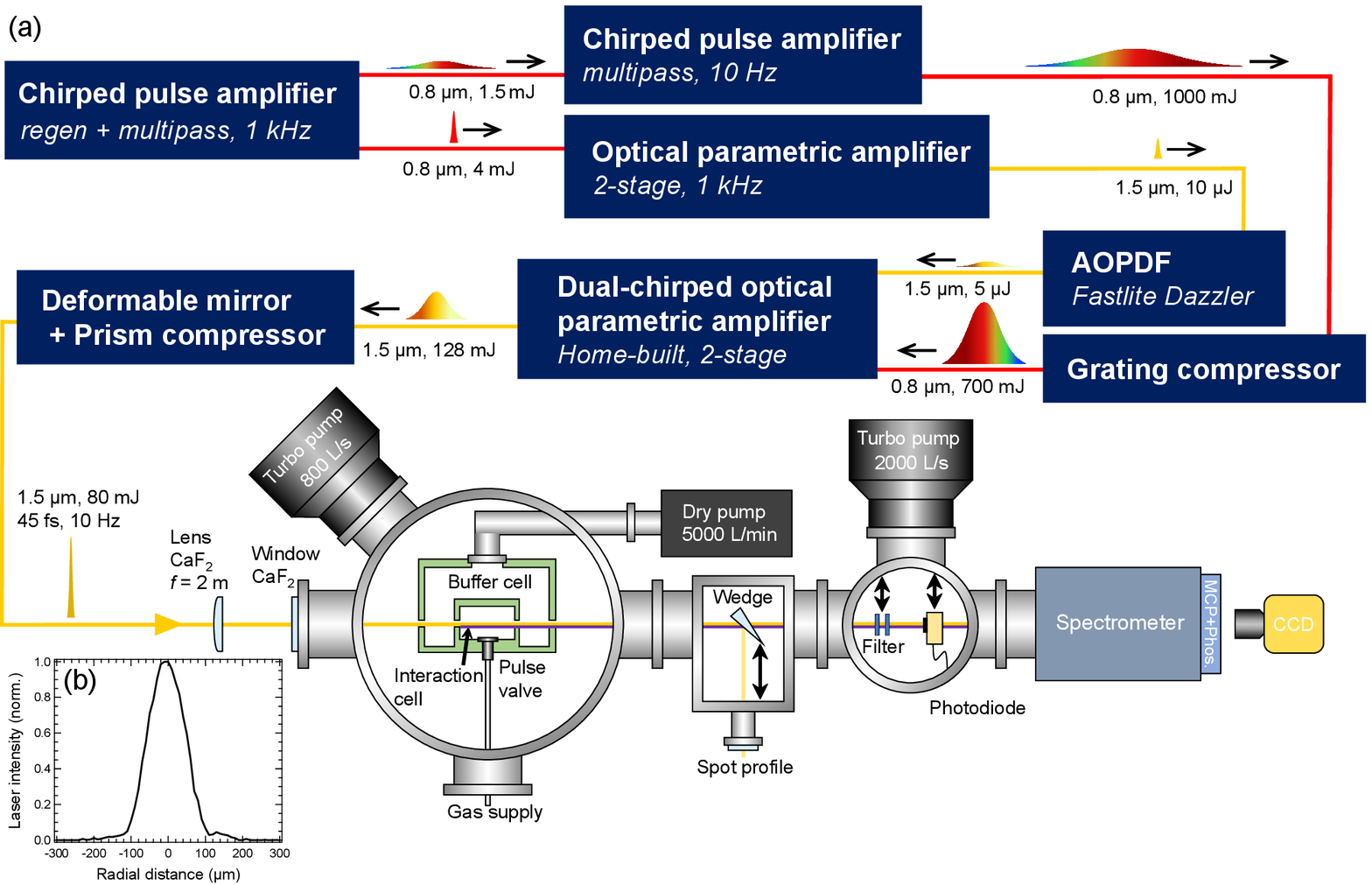}
\caption{\label{fig:overview} (a) Schematic of driving laser system and vacuum chamber for HHG. For the driving laser, 2-stage DC-OPA laser system pumped by a commercial Ti:sapphire laser system was used. HHs were generated in a double-structured pulsed-gas cell and characterized by a photodiode and spectrometer. (b) One-dimensional focusing profile. The beam is focused loosely by the optical lens with the focal length of 2.0 m. 
}
\end{figure*}

An obvious advantage of loose-focusing geometry is lower optimal gas pressure compared to that of the tight-focusing geometry. 
Fig.~\ref{fig:popt} (a) shows phase mismatch as a function of spot diameter and gas pressure for the fundamental wavelength of 1.55 $\rm\mu m$, the HH photon energy of 284 eV, and  the ionization degree of 0.07$\%$ (same as our expected experimental value) for He.
The white curve shows the optimal pressure for phase-matching.
The optimized gas pressure is inversely proportional to the spot area. As the spot diameter increases, phase-matching condition becomes insensitive to the geometric phase difference .
It is important to emphasize that lower gas pressure enables gas consumption reduction and results in vacuum deterioration. 
Low optimized gas pressure requires decrease in the number of HH emitters. However, it can be compensated by increasing the interaction length up the absorption limit. The latter allows maintaining sufficient number of HH emitters\cite{PhysRevLett.82.1668}.

Fig.~\ref{fig:popt} (b) shows the optimal pressure as a function of spot diameter and fundamental wavelength in the cut-off region. 
The ionization degree is set to 0.07$\%$.
An increase in the fundamental wavelength from Ti:sapphire laser to 0.8 $\mathrm{\mu m}$, results in rapid increase of the optimal pressure. 
To mitigate high-pressure effect, loose focusing geometry is suitable for HHG. Consequently, the driving laser should generate longer wavelength.

\section{EXPERIMENT}

\subsection{Setup overview}

Fig.~\ref{fig:overview}  shows the schematic of nJ-class HHG apparatus.  It consists of the driving laser system based on the DC-OPA and the vacuum chambers for HHG. 
DC-OPA is a family of optical parametric chirped-  pulse amplifiers (OPCPA). 
For DC-OPA, we used a chirped femtosecond pumping pulse from  Ti:sapphire laser instead of a picosecond one that is typical for  Yb:YAG lasers.  
The seed pulse has also been stretched to overlap with the pump pulse.
These two chirped pulses bring the name of the technique, dual-chirped optical parametric amplification. 
Our DC-OPA system can generate IR pulses in the range of 1.2 - 4.0 $\mathrm{\mu m}$ with TW-class peak power \cite{7927712, Fu2018}.
Moreover, DC-OPA allows generation of  few-cycle high-energy pulses \cite{Xu:20}.
For HHG, we set the center wavelength of the  driving laser to 1.55 $\mathrm{\mu m}$ that corresponds to a transmission window of the air.
The seed pulses were stretched in time by the acousto-optic programmable dispersive filter (AOPDF).
The chirped seed pulse was amplified by the chirped Ti:sapphire pulse with the type-II BBO crystal in two stages.
After the amplification, the pulse energy reached 130 mJ with the pulse duration of a few picoseconds.
The output pulse was coarsely compressed by the prism compressor (SK1300) and finely compressed by tuning the phase of AOPDF.
The final pulse energy and duration were up to 80 mJ and 45 fs, respectively.

The driving pulse was loosely focused by a 2 m CaF$_{2}$ convex lens and delivered to the target chamber through a CaF$_{2}$  window.
Here, the focusing ability of the driving pulse is a key parameter that affects the efficiency of  HH beams generation \cite{Wang_2018}.
Fig.~\ref{fig:overview} (b) shows the spot profile for  1.55 $\mathrm{\mu m}$ pulse focused by 2.0 m convex lens. 
Measured intensity profile had a gaussian-like distribution with the beam diameter of 190 $\mathrm{\mu m}$.
HH pulse energy can be measured directly using a photodiode (Opto Diode AXUV100G) with proper filters. 
HH spectrum was taken by a 2400 grooves/mm flat-field grating, a microchannel plate with a phosphor screen.
The distance from the gas cell to the slit of the spectrometer was approximately 2 m.

\subsection{\label{sec:33}Double-structured gas cell with 10 Hz operation}
As we explained above, the phase-matching condition must be satisfied to generate intense HHs. 
To meet the absorption limit condition for HHG,  long and a homogeneous gas target is required. 
Generally, a gas jet, a gas cell, and a capillary waveguide have been employed to supply HHG gas media.
The characteristics of each of these gas media are shown in Table \ref{tab:table3}.
Gas jet geometry is the easiest one as it requires no beam alignment.  Moreover, gas consumption is relatively low, so that the vacuum leak is negligible.
However, its typical interaction length is about few millimeters. 
On the other hand,  the capillary waveguide can extend an interaction length.
At the same time, gas density distribution along the  propagation direction is not homogeneous \cite{gebhardt2020bright}.
Thus, the static gas cell is widely adopted for loose-focusing geometry. It allows an easy extension of  the cell length and ensures homogeneous gas density distribution along the laser propagation axis.
However, continuous gas leaking from pinholes breaks the vacuum.
The gas consumption is relatively high, especially when high gas pressure is supplied.
In order to decrease the amount of gas leaking and to realize long and homogeneous high-pressure gas target, we adopted and designed a gas cell with the pulse operation.
Additionally, we made it double-structured for the differential pumping to realize extremely high-pressure operation. 

\begin{table*}
\caption{\label{tab:table3}Comparison of HHG gas targets. }
\begin{ruledtabular}
\begin{tabular}{ccccc}
&Gas jet&Gas cell&Capillary waveguide&Pulsed gas cell\\
\hline
Medium length&a few mm&several cm &several cm&several cm\\
Gas density profile&inhomogeneous&homogeneous&inhomogeneous along the propagation direction&instantaneous homogeneous\\
Repetition& $<$ 1 kHz & continuous & continuous & $<$ 1 kHz\\
Gas consumption&low&high&high&a low\\
Evacuation of a gas medium&easy&difficult&difficult&easy\\
\end{tabular}
\end{ruledtabular}
\end{table*}

\begin{figure}[ht]
\includegraphics[width=\linewidth]{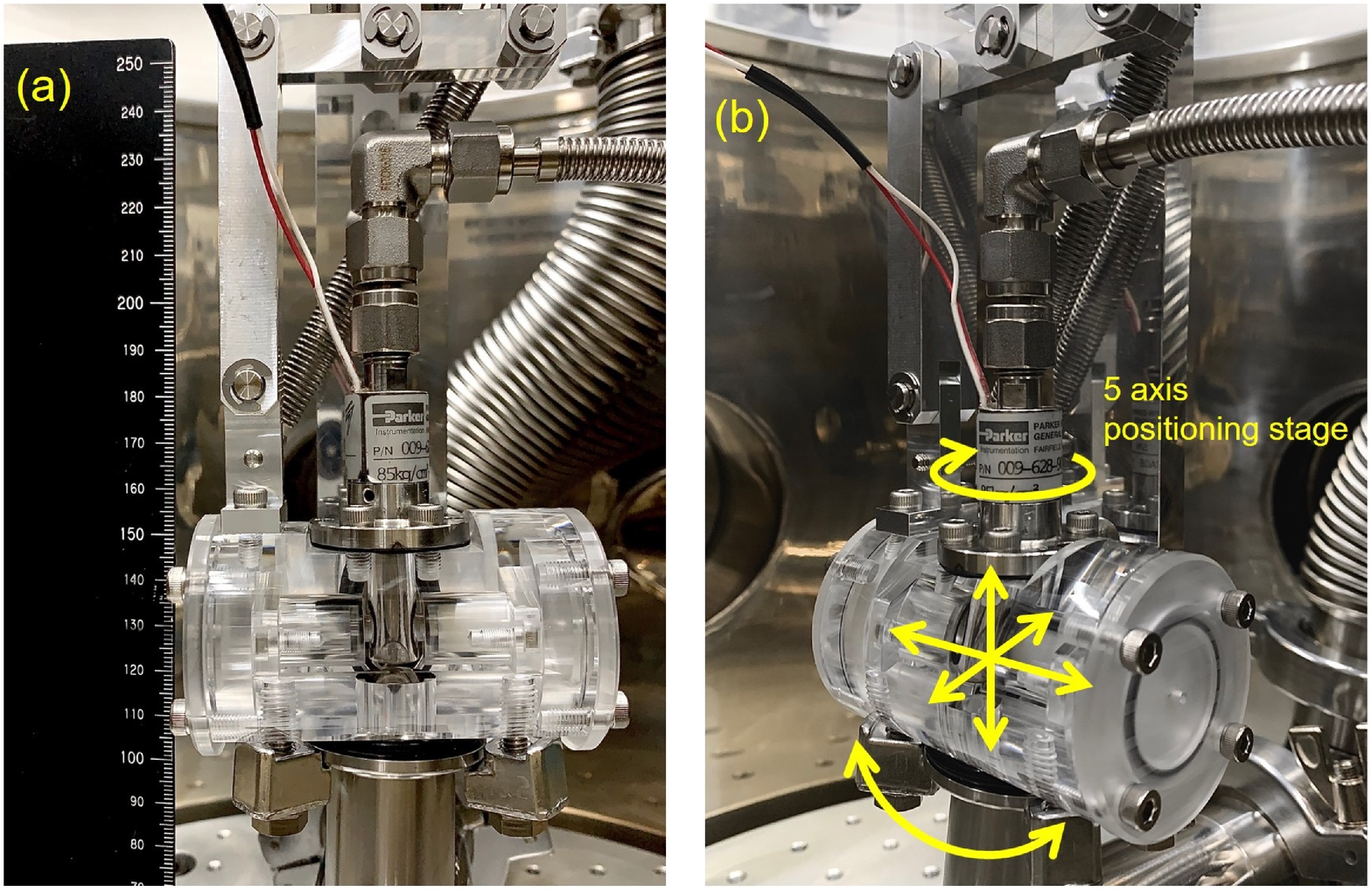}
\includegraphics[width=\linewidth]{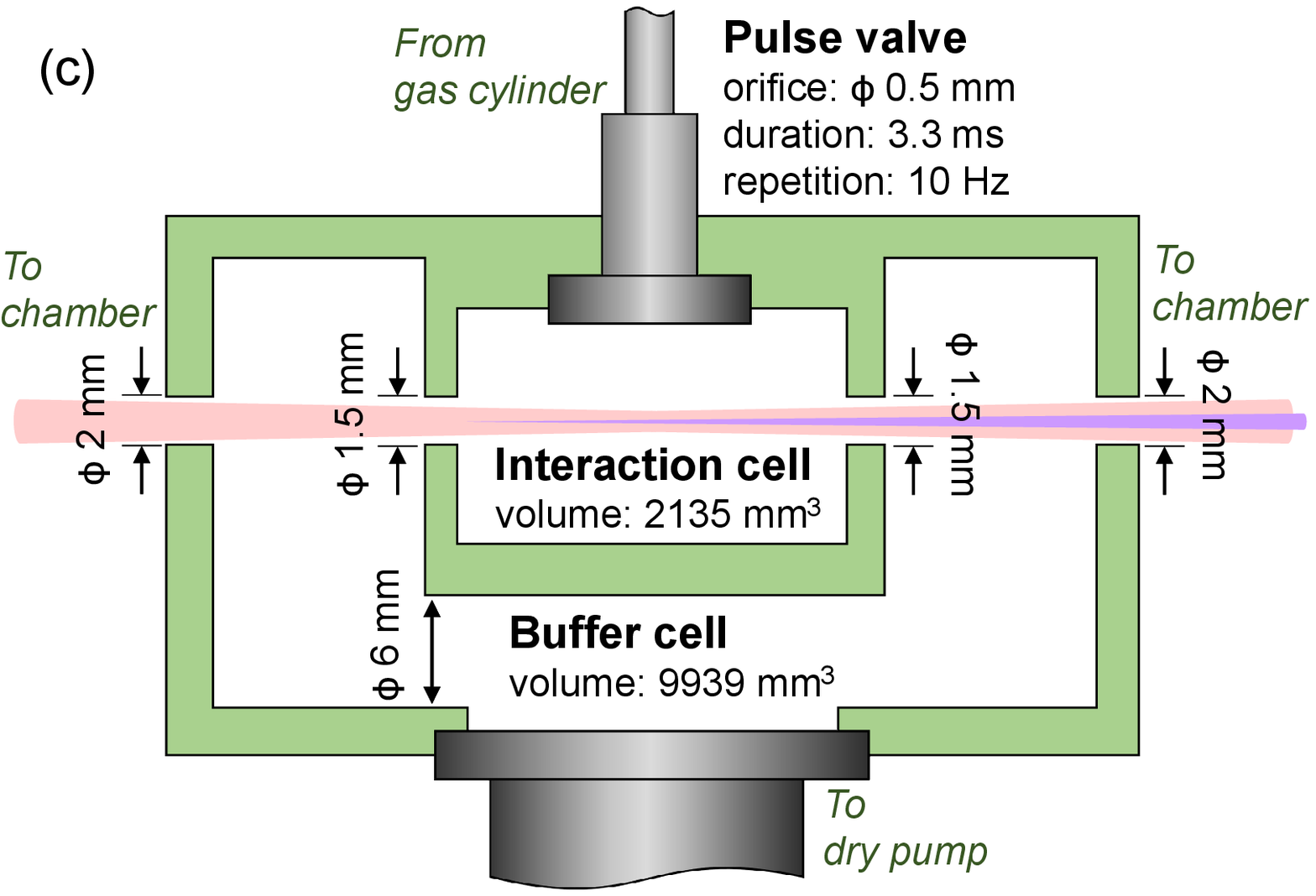}
\caption{(a) Side view of the double-structured pulsed-gas cell with a ruler. 
(b) Image of the double-structured pulsed-gas cell. 
The cell can be aligned along the optical path by using a five-axis positioning stage (XYZ and two rotating axes). 
(c) Schematic of the double-structured pulsed-gas cell. The interaction length is 30 mm. 
The parameters used for the simulation are also shown.
}
\label{fig:gascell}
\end{figure}

Fig.~\ref{fig:gascell} shows the images and schematic of our double-structured pulsed-gas cell. 
The cell consists of a 10 Hz pulse valve (Parker Hannifin 009-628-900), an inner interaction cell, and an outer buffer cell designed as a differential pumping system.
The cell is made of transparent acrylic resin, so that an inside plasma fluorescence is observable. 
The pulse valve opening is synchronized with the 10 Hz driving laser system. 
The target gas inside the inner cell leaks to the outer buffer cell passing through pinholes and then is mostly evacuated by a 5000-L/min dry pump.
The residual gas flows to the vacuum chamber through pinholes in the outer buffer cell and then is exhausted by a 800-L/s turbo molecular pump. 
The four pinholes can be aligned along the optical path by using a five-axis positioning stage ($X-Y-Z$ and two rotating axes). 
In order to avoid energy loss at the pinholes, we set inner and outer pinhole diameters to 1.5 mm and 2mm, respectively. These diameters are  sufficiently large compared to the spot diameter of 190 $\mu$m (see fig.~\ref{fig:overview} (b)).
 
In order to evaluate gas pressure in both inner and outer buffer cells, we simulated gas flow as a function of time by using MATLAB/Simulink. 
The chamber pressure was calculated from the mass flow through two  outer pinholes  using the following differential equation:

\begin{equation}
\label{de}
  \frac{dm(t)}{dt} = \dot{m}_{\mathrm{in}}(t) - \frac{Q_\mathrm{v}}{V}m(t),
\end{equation}
where $m(t)$ is the total mass of the gas in the chamber, $\dot{m}_{\mathrm{in}}(t)$ is the mass flow through two outer pinholes, $Q_\mathrm{v}$ is the volume flow rate of the turbo molecular pump, and $V$ is the volume of the chamber. 
For the sake of simplicity, we assumed that $Q_\mathrm{v}$ was constant. 
The solution of equation (\ref{de}) reads as

\begin{equation}
\label{sol}
 m(t) = \exp\left(\int \frac{Q_\mathrm{v}}{V} \mathrm{dt}\right) \left[\int  \exp\left(\frac{Q_\mathrm{v}}{V}\right) \dot{m}_{\mathrm{in}}(t) \mathrm{dt}\right].
\end{equation}
Equation (\ref{sol}) can be used to determine chamber pressure as a function of time.

\begin{figure}[htbp]
\includegraphics[width=\linewidth]{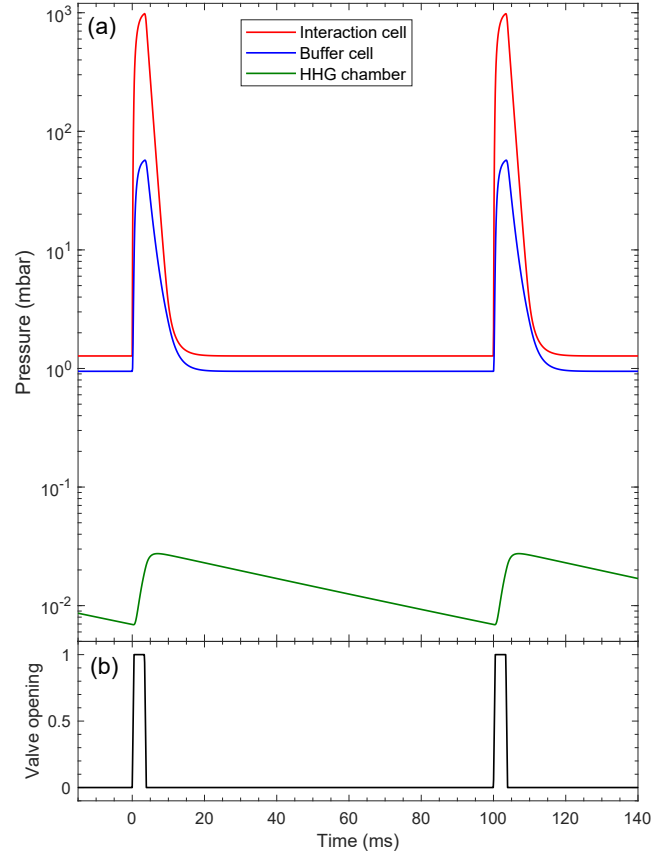}
\caption{\label{fig:timepressure} (a) Calculated pressures in double-structured pulsed-gas cell and HHG chamber as a function of time. The gas medium is He and its backing pressure is 10 bar. (b) Timing of the pulse valve opening. The values of 0 and 1 correspond to close and open valve, respectively. The opening time and repetition rate are 3.3 ms and 10 Hz, respectively.
}
\end{figure}

Fig.~\ref{fig:timepressure} shows the calculation results for the time-dependent gas pressures in the interaction cell, the buffer cell, and the chamber. 
In this calculation, we used the design parameters  shown in fig.~\ref{fig:gascell} (c). 
The gas medium was He and its backing pressure was fixed at 10 bar.
The valve was opened for 3.3 ms and the time between close and open states was assumed to be 0.33 ms.
We also assumed that when the valve was closed, the leak flow from the valve existed and was equal to 1/1000 of the flow from the open valve.
For the sake of simplicity, the gas medium in the cell or chamber was assumed to diffuse instantaneously.
As a result, the calculated pressure corresponded to the spatial mean pressure at equilibrium in the cell or chamber.
%This approximation is accurate in this time range because the diffusion rate of He atom is as fast as 1.4 m/ms.
As the pulse valve was opened, the pressure in all chambers increased rapidly.
When the pulse valve was closed, the pressure in the interaction cell reached its maximum, 10 $\%$ of the backing pressure.
Upon certain time elapsed after valve was closed , the pressures in the gas cell converged to a constant value as the balance between the evacuation and the leak flows from the valve was achieved.
Time-dependent pressure curves were similar to those shown in fig.~\ref{fig:timepressure} for the backing pressure ranging from 0.5 bar to 20 bar.
The maximum gas pressure  in the buffer cell was only $6\%$ of the maximum gas pressure in the interaction cell.
The large difference between the gas pressure in each region can suppress the driving laser distortion and the absorption of HH beams, except the interaction region.
The chamber pressure was low enough to achieve long-time operation of a turbo molecular pump. The latter was crucial  in realization of ideal vacuum condition desirable for practical applications.
Compared to continuum-flow operation, gas consumption was reduced by to ${96.45\%}$.  
This consumption was low enough for daily continuous operations of HHG at reasonable cost (a 0.047-$\mathrm m^3$ 118-bar He gas cylinder per month on condition of 5-hour workload per day with the backing pressure of 10 bar).

\begin{figure*}[htbp]
\includegraphics[width=\linewidth]{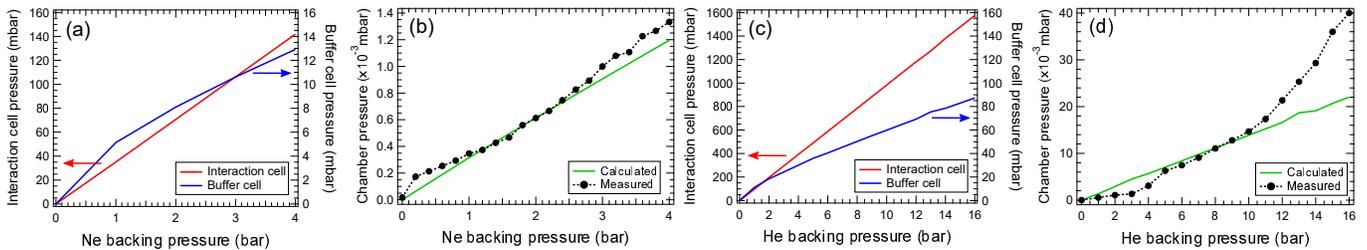}
\caption{\label{fig:pressures} Pressure in double-structured pulsed-gas cell as a function of backing pressure. (a)(c) Calculated pressure in interaction and buffer cells as a function of the backing pressure of Ne (a) and He (c). Each calculated point corresponds to the highest pressure during 100 ms time interval. (b)(d) Calculated and measured pressure in chamber as a function of backing pressure of Ne (b) and He (d). Each calculated point corresponds to the mean pressure during 100 ms time interval.}
\end{figure*}

Fig.~\ref{fig:pressures} (a) and (c) show calculated pressures in pulsed-gas cell as a function of the backing pressure in Ne and He, respectively.
Each point of the pressure in interaction  and buffer cells corresponded to the maximum value (see fig.~\ref{fig:gascell}). 
The pressure in the interaction cell is proportional to the backing pressure. 
Ne and He pressure in the interaction cell were $3.55\%$ and $9.85\%$ of the backing pressure, respectively. 
The buffer pressure of Ne and He were $9-15\%$ and $5-10\%$ of the pressure in the interaction cell, respectively. 
The differential pumping system was capable of reducing   buffer pressure 10-fold  compared to interaction cell pressure and suppressed the reabsorption of HHs.

Fig.~\ref{fig:pressures} (b) and (d) show calculated and measured chamber pressures with respect to the backing pressure in Ne and He. 
Each point of calculated pressure in the chamber corresponded to the mean value
during 100 ms time interval (see fig.~\ref{fig:gascell}). 
In  case of Ne, calculated and measured data were in good agreement. The latter can be due to relatively low absolute values that were insufficient to achieve long-time operation of a turbo molecular pump. 
In case of He,  calculated data  also  agreed well with the measured one below 10 bar. 
We assumed that the volume flow rate of the turbo molecular pump, $Q_\mathrm{v}$, was constant.
However, in practice, the pumping flow rate gradually decreases above $10^{-1}$ mbar pressure. 
As a result, the measured chamber pressure becomes higher than the calculated one when the backing pressure exceeds 10 bar. 
For both Ne and He cases, the pressure in spectrometer was less than $3\times10^{-6}$ mbar, that was low enough for MCP operation.

\section{RESULTS}
\subsection{\label{sec:41}nJ-class HHG in the water-window region}
\begin{figure}[ht]
\includegraphics[width=0.9\linewidth]{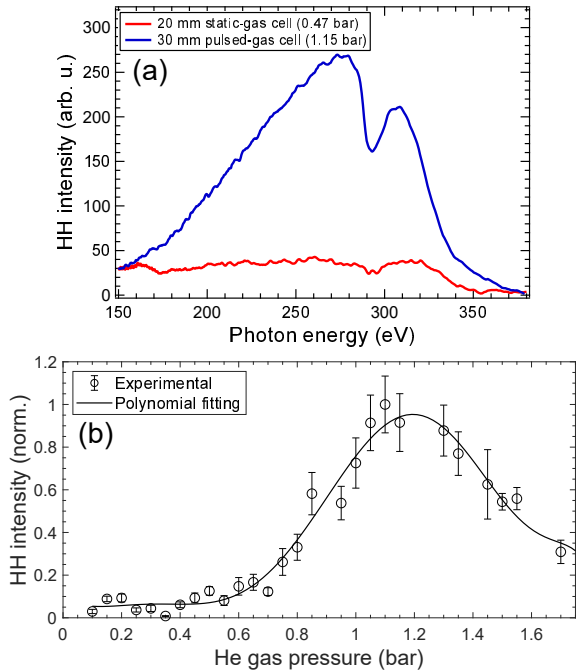}
\caption{\label{fig:HePD} (a) Water-window HH spectra generated in static-gas cell (red curve) and pulsed-gas cell (blue curve). The pressure in static cell is limited by the vacuum degree in the HHG chamber. (b) He pressure dependence from HH intensity for pulsed-gas cell. HH intensity reached the peak value at around 1.2 bar (12 bar backing pressure). The limiting static-gas-cell pressure of 0.47 bar was far from the peak. Subfigure (b) is reproduced from \cite{Fu2020} with permission.}
\end{figure}

To demonstrate an intense high-order harmonic generation in the water-window region, we supplied double-structured pulsed-gas cell with the high-pressure He. 
We set the center wavelength of the driving laser to 1.55 $\mathrm{\mu m}$ and its pulse duration to 45 fs.
Fig.~\ref{fig:HePD} (a) shows the comparison of the HH spectra generated in double-structured static-gas cell (red curve) and double-structured pulsed-gas cell (blue curve). 
The cut-off energy was approximately 360 eV, which entered into the water-window region.
The dip at 284 eV might be caused by the carbon contamination on the grating surface of the spectrometer.
When the static-gas cell was used, the IR pulse energy and the interaction length were set to  42 mJ  and 20 mm, respectively.
The highest possible gas pressure was 0.47 bar.
When the gas pressure surpassed 0.47 bar, turbo molecular pump was unable to operate due to the vacuum leakage.
The gas consumption was also high enough and reached 0.010-$m^3$ 118-bar He gas cylinder per hour.
In order to increase gas pressure and reduce vacuum leakage, we employed double-structured pulsed-gas cell.
We successfully increased HH intensity by 10 times using pulsed-gas operation due to increase in gas pressure.
Fig.~\ref{fig:HePD} (b) shows He pressure dependence from HH intensity for the pulsed-gas cell.
The HH intensity reached the peak value at around 1.2 bar (12 bar backing pressure) and its spectrum corresponded to the blue curve in fig.~\ref{fig:HePD} (a).
It means that the phase-matching condition is optimized at 1.2 bar and the limit of  static-gas-cell pressure (0.47 bar) is far from the optimal condition.
The total HH pulse energy in the water-window region was measured with a photodiode. It amounted to  3.53 nJ with the conversion efficiency of $7.20\times 10^{-8}$.
The conversion efficiency was improved by more than 10 times compared to previously reported data.
A possible reason for such a dramatic change in conversion efficiency was that we used different phase-matching condition. In particular,  it was different from the transient phase-matching under the tight-focusing geometry and more stable along the propagation direction.  The latter was attributed to the fact that  the neutral gas medium almost didn't modulate the driving laser field\cite{Lai:11}.
Robust phase-matching condition enabled accumulation of HHs by increasing the interaction length.  Consequently, the conversion efficiency was also improved.
Moreover, TW-class IR laser system  allowed for  a 100-fold more intense soft x-ray harmonic beam generation (compared to previous studies) in the water-window region.

\subsection{\label{sec:43}Absorption spectroscopy analysis near carbon K-edge}

\begin{figure}[ht]
\includegraphics[width=\linewidth]{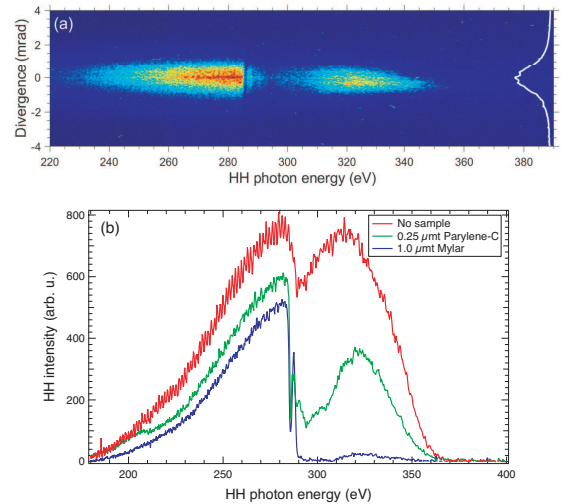}
\caption{\label{fig:HeHHG} (a) Two-dimensional HH spectrum from He generated after passing 0.25-$\mathrm{\mu m}$ Parylene-C thin film. The white curve shows one-dimensional profile of total HHs. (b) HH spectrum from He driven by 1.55-$\mathrm{\mu m}$ 45-fs 49-mJ pulses (red curve) and HH absorption spectra after 0.25-$\mathrm{\mu m}$ Parylene-C (green curve) and a 1.0-$\mathrm{\mu m}$ Mylar (blue curve) thin films. The cut-off energy was approximately 360 eV. The fine absorption structures were clearly observed near carbon K-edge  (absorption dip around 284 eV). The dip of the red curve was caused by the contamination in the chambers or on the grating.}
\end{figure}

Here we present the absorption spectroscopy analysis near carbon K-edge using the light source described above.
In order to improve spectral resolution, we have upgraded spectrometer's  MCP with  the 16-bit x-ray CCD camera. 
The red curve in fig.~\ref{fig:HeHHG} (b) is the original HH spectrum of He. 
The green and the blue curves in fig.~\ref{fig:HeHHG} (b) are HH absorption spectra after  0.25-$\mathrm{\mu m}$ Parylene-C and 1.0-$\mathrm{\mu m}$ Mylar thin films, respectively.
Fine absorption structures were clearly observed near carbon K-edge. 
Fig.~\ref{fig:HeHHG} (a) shows two-dimensional HH spectrogram after 0.25-$\mathrm{\mu m}$ Parylene-C thin film. 
The HHs were bright enough to obtain the fine absorption structure with spatial information.
To illustrate a simple application, let us consider  lower part of the HHs only passing thin film sample. In this case, we can simultaneously get reference HH spectrum from the upper part and calculate the absorbance of the sample even from a single shot.
The detailed discussion of fine absorption structures can be found  in our previous study \cite{Fu2020}.

\section{FUTURE PROSPECTS}
In order to achieve attosecond temporal resolution, an isolated attosecond pulse (IAP) is required instead of the attosecond pulse train. 
Supercontinuum HHs were generated using few-cycle IR pulses while 53-attosecond isolated pulse can be obtained from Ne in the water-window region with OPA laser system\cite{Li2017}.
Recently our DC-OPA system has been improved to generate 1.7-$\mathrm{\mu m}$ two-cycle TW-class pulses\cite{Xu:20,Xu:2020}. These researchers obtained similar pulse duration and  OPA laser spectrum \cite{Li2017}.
The conversion efficiency of water-window IAP under our apparatus can be expected by scaling the conversion efficiency of our Ne HHG \cite{Fu2020} with $\propto \lambda^{-6}$.
The measured conversion efficiency was $7.4 \times 10^{-9}$ at 250 eV in 1$\%$ bandwidth while the scaling factor was 0.57.
Therefore, the estimated conversion efficiency is $4.3 \times 10^{-9}$ at 250 eV in 1$\%$ bandwidth.
The estimated total HH pulse energy is approximately 2 nJ with tens of attosecond pulse duration.
The optimal pressure for the phase-matching is 188 mbar and can be achieved for the backing pressure of 5.3 bar in the double-structured pulsed-gas cell.
This intense water-window IAP has the potential to open the door for unexplored attosecond science in the water-window region including intra-atomic energy transfer \cite{Sisourat2010}, ultrafast electron transfer\cite{Fohlisch2005}, and so on.

\section{SURMARY}
\begin{figure}[ht]
\includegraphics[width=\linewidth]{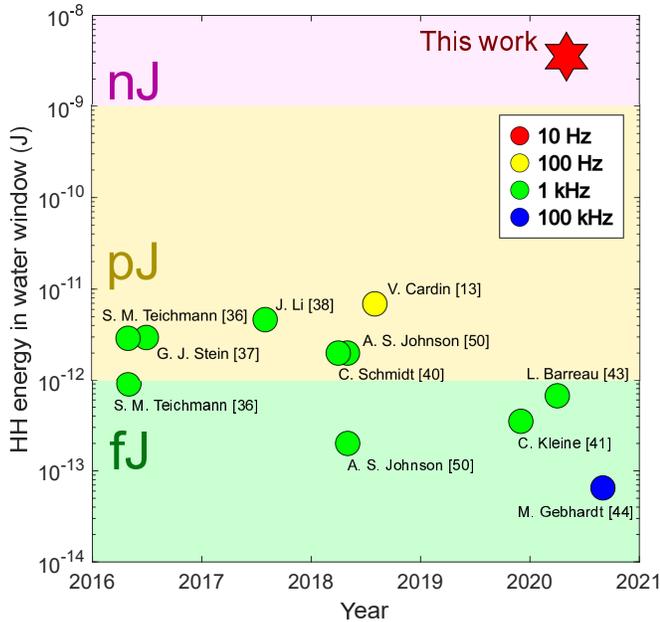}
\caption{\label{fig:Summary}Reported total HH pulse energies in the water-window region in the recent 5 years. The color of the points corresponds to the repetition frequency.}
\end{figure}

We have developed an apparatus for nJ-class HHG in the water-window region. 
Our TW-class OPA laser system is capable of generating the first HHG in a-few-meter focusing geometry.
Despite  the optimal gas pressure for the phase-matching in loose-focusing geometry being relatively low, the leak flow from the gas cell seriously deteriorated the surrounding vacuum. As a result,  the gas consumption was extremely high when the gas was continuously supplied.
Therefore, we designed and fabricated double-structured pulsed-gas cell, simulated and measured the actual gas pressure in it.
The calculated pressure was in good agreement with the measured value.
The gas cell developed in this study can dramatically reduce both vacuum leakage and gas consumption.
Further optimization can be achieved by increasing He pressure in the interaction region to over 1 bar with 1.5-mm-diameter pinholes. As a result, gas consumption can be reduced to 3.55$\%$ of the continuum flow operation.
High-pressure gas increases water-window HH intensity by 10 times compared to the that for the continuum-gas cell.
Total HH pulse energy in the water-window region amounted to 3.53 nJ, a more than 100-fold increase compared to previously reported values.
Fig.~\ref{fig:Summary} (a) shows the comparison of the reported total HH pulse energies in the water-window region in the recent 5 years.
Up to date, HHG apparatus developed in this study, is the only one that can reach  nJ-class pulse energy.
As a first application of this light source, we demonstrated the absorption spectroscopy analysis near carbon K-edge.
Two-dimensional absorption spectrogram has demonstrated clear fine absorption structures.
In future, our robust method for water-window HHG is expected to have energy-scalability and will realize the single-shot transient absorption spectroscopy or coherent diffractive imaging in the water-window region. Based on our apparatus, we expect obtaining nJ isolated attosecond pulses in the water-window region in the nearest future.

\begin{acknowledgments}
This work was supported in part by the Ministry
of Education, Culture, Sports, Science and Technology
of Japan (MEXT) through Grant-in-Aid under Grant
17H01067, in part by the MEXT Quantum Leap Flagship
Program (Q-LEAP) Grant Number JP-MXS0118068681,
in part by the FY 2019 President discretionary funds of
RIKEN, in part by the Matsuo Foundation 2018, and in part by RIKEN Junior Research Associate Program. 
K. M. gratefully acknowledges financial support from Grantsin-Aid for Scientific Research 19H0562.
Y. F. gratefully acknowledges financial support from Major Science and Technology Infrastructure Pre-research Program of CAS (J20-021-I\hspace{-.1em}I\hspace{-.1em}I), Key Deployment Research Program of XIOPM (No.S19-020-I\hspace{-.1em}I\hspace{-.1em}I), Natural Science Basic Research Program of Shaanxi (No.2019JCW-03), National Natural Science Foundation of China (No.61690222).
\end{acknowledgments}

%\nocite{*}
%\%bibliography{aipsamp}% Produces the bibliography via BibTeX.
%%\bibliographystyle{unsrt}

\providecommand{\noopsort}[1]{}\providecommand{\singleletter}[1]{#1}%


\begin{thebibliography}{10}

\bibitem{PhysRevX.7.041030}
Seth~L. Cousin, Nicola Di~Palo, B\'{a}rbara Buades, Stephan~M. Teichmann,
  M.~Reduzzi, M.~Devetta, A.~Kheifets, G.~Sansone, and Jens Biegert.
\newblock Attosecond streaking in the water window: A new regime of attosecond
  pulse characterization.
\newblock {\em Phys. Rev. X}, 7:041030, Nov 2017.

\bibitem{Saito2016}
Nariyuki Saito, Nobuhisa Ishii, Teruto Kanai, Shuntaro Watanabe, and Jiro
  Itatani.
\newblock Attosecond streaking measurement of extreme ultraviolet pulses using
  a long-wavelength electric field.
\newblock {\em Scientific Reports}, 6(1):35594, Oct 2016.

\bibitem{Gaumnitz:17}
Thomas Gaumnitz, Arohi Jain, Yoann Pertot, Martin Huppert, Inga Jordan,
  Fernando Ardana-Lamas, and Hans~Jakob W\"{o}rner.
\newblock Streaking of 43-attosecond soft-x-ray pulses generated by a passively
  {CEP}-stable mid-infrared driver.
\newblock {\em Opt. Express}, 25(22):27506--27518, Oct 2017.

\bibitem{Zhang:04}
X.~Zhang, A.~R. Libertun, A.~Paul, E.~Gagnon, S.~Backus, I.~P. Christov, M.~M.
  Murnane, H.~C. Kapteyn, R.~A. Bartels, Y.~Liu, and D.~T. Attwood.
\newblock Highly coherent light at 13 nm generated by use of
  quasi-phase-matched high-harmonic generation.
\newblock {\em Opt. Lett.}, 29(12):1357--1359, Jun 2004.

\bibitem{doi:10.1063/1.3475689}
M.~Fie\ss, M.~Schultze, E.~Goulielmakis, B.~Dennhardt, J.~Gagnon,
  M.~Hofstetter, R.~Kienberger, and F.~Krausz.
\newblock Versatile apparatus for attosecond metrology and spectroscopy.
\newblock {\em Review of Scientific Instruments}, 81(9):093103, 2010.

\bibitem{Chini:09}
Michael Chini, Hiroki Mashiko, He~Wang, Shouyuan Chen, Chenxia Yun, Shane
  Scott, Steve Gilbertson, and Zenghu Chang.
\newblock Delay control in attosecond pump-probe experiments.
\newblock {\em Opt. Express}, 17(24):21459--21464, Nov 2009.

\bibitem{Mashiko:10}
Hiroki Mashiko, M.~Justine Bell, Annelise~R. Beck, Mark~J. Abel, Philip~M.
  Nagel, Colby~P. Steiner, Joseph Robinson, Daniel~M. Neumark, and Stephen~R.
  Leone.
\newblock Tunable frequency-controlled isolated attosecond pulses characterized
  by either 750 nm or 400 nm wavelength streak fields.
\newblock {\em Opt. Express}, 18(25):25887--25895, Dec 2010.

\bibitem{McPherson:87}
A.~McPherson, G.~Gibson, H.~Jara, U.~Johann, T.~S. Luk, I.~A. McIntyre,
  K.~Boyer, and C.~K. Rhodes.
\newblock Studies of multiphoton production of vacuum-ultraviolet radiation in
  the rare gases.
\newblock {\em J. Opt. Soc. Am. B}, 4(4):595--601, Apr 1987.

\bibitem{Ferray_1988}
M~Ferray, A~L'Huillier, X~F Li, L~A Lompre, G~Mainfray, and C~Manus.
\newblock Multiple-harmonic conversion of 1064 nm radiation in rare gases.
\newblock {\em Journal of Physics B: Atomic, Molecular and Optical Physics},
  21(3):L31--L35, Feb 1988.

\bibitem{Takahashi:02}
Eiji Takahashi, Yasuo Nabekawa, and Katsumi Midorikawa.
\newblock Generation of 10-{$\mathrm{\mu}$J} coherent extreme-ultraviolet light
  by use of high-order harmonics.
\newblock {\em Opt. Lett.}, 27(21):1920--1922, Nov 2002.

\bibitem{Takahashi2013}
Eiji~J. Takahashi, Pengfei Lan, Oliver~D. M\"{u}cke, Yasuo Nabekawa, and
  Katsumi Midorikawa.
\newblock Attosecond nonlinear optics using gigawatt-scale isolated attosecond
  pulses.
\newblock {\em Nature Communications}, 4(1):2691, Oct 2013.

\bibitem{PhysRevLett.101.253901}
Eiji~J. Takahashi, Tsuneto Kanai, Kenichi~L. Ishikawa, Yasuo Nabekawa, and
  Katsumi Midorikawa.
\newblock {Coherent Water Window X Ray by Phase-Matched High-Order Harmonic
  Generation in Neutral Media}.
\newblock {\em Phys. Rev. Lett.}, 101:253901, Dec 2008.

\bibitem{Cardin_2018}
V~Cardin, B~E Schimdt, N~Thir{\'{e}}, S~Beaulieu, V~Wanie, M~Negro, C~Vozzi,
  V~Tosa, and F~L{\'{e}}gar{\'{e}}.
\newblock Self-channelled high harmonic generation of water window soft x-rays.
\newblock {\em Journal of Physics B: Atomic, Molecular and Optical Physics},
  51(17):174004, Aug 2018.

\bibitem{Pupeikis:20}
J.~Pupeikis, P.-A. Chevreuil, N.~Bigler, L.~Gallmann, C.~R. Phillips, and
  U.~Keller.
\newblock Water window soft x-ray source enabled by a 25 {W} few-cycle 2.2
  {$\mu$ m} {OPCPA} at 100 {kHz}.
\newblock {\em Optica}, 7(2):168--171, Feb 2020.

\bibitem{Takahashi:03}
Eiji Takahashi, Yasuo Nabekawa, Muhammad Nurhuda, and Katsumi Midorikawa.
\newblock Generation of high-energy high-order harmonics by use of a long
  interaction medium.
\newblock {\em J. Opt. Soc. Am. B}, 20(1):158--165, Jan 2003.

\bibitem{PhysRevA.68.023808}
Eiji Takahashi, Valer Tosa, Yasuo Nabekawa, and Katsumi Midorikawa.
\newblock Experimental and theoretical analyses of a correlation between
  pump-pulse propagation and harmonic yield in a long-interaction medium.
\newblock {\em Phys. Rev. A}, 68:023808, Aug 2003.

\bibitem{doi:10.1063/1.1637949}
Eiji~J. Takahashi, Yasuo Nabekawa, and Katsumi Midorikawa.
\newblock Low-divergence coherent soft x-ray source at 13 nm by high-order
  harmonics.
\newblock {\em Applied Physics Letters}, 84(1):4--6, 2004.

\bibitem{PhysRevA.84.053819}
W.~Boutu, T.~Auguste, J.~P. Caumes, H.~Merdji, and B.~Carr\'{e}.
\newblock Scaling of the generation of high-order harmonics in large gas media
  with focal length.
\newblock {\em Phys. Rev. A}, 84:053819, Nov 2011.

\bibitem{Heyl_2012}
C~M Heyl, J~G\"{u}dde, A~L'Huillier, and U~H\"{o}fer.
\newblock High-order harmonic generation with {$\mu$J} laser pulses at high
  repetition rates.
\newblock {\em Journal of Physics B: Atomic, Molecular and Optical Physics},
  45(7):074020, Mar 2012.

\bibitem{doi:10.1063/1.4812266}
P.~Rudawski, C.~M. Heyl, F.~Brizuela, J.~Schwenke, A.~Persson, E.~Mansten,
  R.~Rakowski, L.~Rading, F.~Campi, B.~Kim, P.~Johnsson, and A.~L’Huillier.
\newblock A high-flux high-order harmonic source.
\newblock {\em Review of Scientific Instruments}, 84(7):073103, 2013.

\bibitem{Heyl_2016}
C~M Heyl, C~L Arnold, A~Couairon, and A~L'Huillier.
\newblock Introduction to macroscopic power scaling principles for high-order
  harmonic generation.
\newblock {\em Journal of Physics B: Atomic, Molecular and Optical Physics},
  50(1):013001, Dec 2016.

\bibitem{Wang_2018}
Yang Wang, Tianyi Guo, Jialin Li, Jian Zhao, Yanchun Yin, Xiaoming Ren, Jie Li,
  Yi~Wu, Matthew Weidman, Zenghu Chang, Marieke~F Jager, Christopher~J Kaplan,
  Romain Geneaux, Christian Ott, Daniel~M Neumark, and Stephen~R Leone.
\newblock Enhanced high-order harmonic generation driven by a wavefront
  corrected high-energy laser.
\newblock {\em Journal of Physics B: Atomic, Molecular and Optical Physics},
  51(13):134005, Jun 2018.

\bibitem{PhysRevA.98.023426}
A.~Nayak, I.~Orfanos, I.~Makos, M.~Dumergue, S.~K\"{u}hn, E.~Skantzakis,
  B.~Bodi, K.~Varju, C.~Kalpouzos, H.~I.~B. Banks, A.~Emmanouilidou,
  D.~Charalambidis, and P.~Tzallas.
\newblock Multiple ionization of argon via multi-{XUV}-photon absorption
  induced by 20-{GW} high-order harmonic laser pulses.
\newblock {\em Phys. Rev. A}, 98:023426, Aug 2018.

\bibitem{Kov_cs_2019}
K~Kov{\'{a}}cs, B~Major, E~Balogh, Cs~P K{\H{o}}r\"{o}s, P~Rudawski, C~M Heyl,
  P~Johnsson, C~L Arnold, A~L'Huillier, V~Tosa, and K~Varj{\'{u}}.
\newblock Multi-parameter optimization of a loose focusing high flux
  high-harmonic beamline.
\newblock {\em Journal of Physics B: Atomic, Molecular and Optical Physics},
  52(5):055402, Feb 2019.

\bibitem{Rupp2017}
Daniela Rupp, Nils Monserud, Bruno Langbehn, Mario Sauppe, Julian Zimmermann,
  Yevheniy Ovcharenko, Thomas M\"{o}ller, Fabio Frassetto, Luca Poletto, Andrea
  Trabattoni, Francesca Calegari, Mauro Nisoli, Katharina Sander, Christian
  Peltz, Marc J.~Vrakking, Thomas Fennel, and Arnaud Rouz\'{e}e.
\newblock Coherent diffractive imaging of single helium nanodroplets with a
  high harmonic generation source.
\newblock {\em Nature Communications}, 8(1):493, Sep 2017.

\bibitem{Baksheaaz3025}
Peter~D. Baksh, Michal Ostr~\v{c}il, Magdalena Miszczak, Charles Pooley,
  Richard~T. Chapman, Adam~S. Wyatt, Emma Springate, John~E. Chad, Katrin
  Deinhardt, Jeremy~G. Frey, and William~S. Brocklesby.
\newblock Quantitative and correlative extreme ultraviolet coherent imaging of
  mouse hippocampal neurons at high resolution.
\newblock {\em Science Advances}, 6(18), 2020.

\bibitem{PhysRevLett.68.3535}
Jeffrey~L. Krause, Kenneth~J. Schafer, and Kenneth~C. Kulander.
\newblock High-order harmonic generation from atoms and ions in the high
  intensity regime.
\newblock {\em Phys. Rev. Lett.}, 68:3535--3538, Jun 1992.

\bibitem{Popmintchev10516}
Tenio Popmintchev, Ming-Chang Chen, Alon Bahabad, Michael Gerrity, Pavel
  Sidorenko, Oren Cohen, Ivan~P. Christov, Margaret~M. Murnane, and Henry~C.
  Kapteyn.
\newblock Phase matching of high harmonic generation in the soft and hard
  {X}-ray regions of the spectrum.
\newblock {\em Proceedings of the National Academy of Sciences},
  106(26):10516--10521, 2009.

\bibitem{Xiong:09}
Hui Xiong, Han Xu, Yuxi Fu, Jinping Yao, Bin Zeng, Wei Chu, Ya~Cheng, Zhizhan
  Xu, Eiji~J. Takahashi, Katsumi Midorikawa, X.~Liu, and J.~Chen.
\newblock Generation of a coherent x ray in the water window region at 1 k{H}z
  repetition rate using a mid-infrared pump source.
\newblock {\em Opt. Lett.}, 34(11):1747--1749, Jun 2009.

\bibitem{PhysRevLett.105.173901}
M.-C. Chen, P.~Arpin, T.~Popmintchev, M.~Gerrity, B.~Zhang, M.~Seaberg,
  D.~Popmintchev, M.~M. Murnane, and H.~C. Kapteyn.
\newblock {Bright, Coherent, Ultrafast Soft X-Ray Harmonics Spanning the Water
  Window from a Tabletop Light Source}.
\newblock {\em Phys. Rev. Lett.}, 105:173901, Oct 2010.

\bibitem{Takahashi2010}
E.~J. Takahashi, T.~Kanai, and K.~Midorikawa.
\newblock High-order harmonic generation by an ultrafast infrared pulse.
\newblock {\em Applied Physics B}, 100(1):29--41, Jul 2010.

\bibitem{Popmintchev1287}
Tenio Popmintchev, Ming-Chang Chen, Dimitar Popmintchev, Paul Arpin, Susannah
  Brown, Skirmantas Ali\v{s}auskas, Giedrius Andriukaitis, Tadas Bal\v{c}iunas,
  Oliver~D. M\"{u}cke, Audrius Pugzlys, Andrius Baltu\v{s}ka, Bonggu Shim,
  Samuel~E. Schrauth, Alexander Gaeta, Carlos Hern\'{a}ndez-Garc\'{i}a, Luis
  Plaja, Andreas Becker, Agnieszka Jaron-Becker, Margaret~M. Murnane, and
  Henry~C. Kapteyn.
\newblock {Bright Coherent Ultrahigh Harmonics in the keV X-ray Regime from
  Mid-Infrared Femtosecond Lasers}.
\newblock {\em Science}, 336(6086):1287--1291, 2012.

\bibitem{Ishii2014}
Nobuhisa Ishii, Keisuke Kaneshima, Kenta Kitano, Teruto Kanai, Shuntaro
  Watanabe, and Jiro Itatani.
\newblock Carrier-envelope phase-dependent high harmonic generation in the
  water window using few-cycle infrared pulses.
\newblock {\em Nature Communications}, 5(1):3331, Feb 2014.

\bibitem{Cousin:14}
S.~L. Cousin, F.~Silva, S.~Teichmann, M.~Hemmer, B.~Buades, and J.~Biegert.
\newblock High-flux table-top soft x-ray source driven by sub-2-cycle, {CEP}
  stable, 1.85-{$\mu$m} 1-k{H}z pulses for carbon {K}-edge spectroscopy.
\newblock {\em Opt. Lett.}, 39(18):5383--5386, Sep 2014.

\bibitem{Silva2015}
Francisco Silva, Stephan~M. Teichmann, Seth~L. Cousin, Michael Hemmer, and Jens
  Biegert.
\newblock Spatiotemporal isolation of attosecond soft {X}-ray pulses in the
  water window.
\newblock {\em Nature Communications}, 6(1):6611, Mar 2015.

\bibitem{Teichmann2016}
S.~M. Teichmann, F.~Silva, S.~L. Cousin, M.~Hemmer, and J.~Biegert.
\newblock {0.5-keV Soft X-ray attosecond continua}.
\newblock {\em Nature Communications}, 7(1):11493, May 2016.

\bibitem{Stein_2016}
Gregory~J Stein, Phillip~D Keathley, Peter Krogen, Houkun Liang, Jonathas~P
  Siqueira, Chun-Lin Chang, Chien-Jen Lai, Kyung-Han Hong, Guillaume~M Laurent,
  and Franz~X K\"{a}rtner.
\newblock Water-window soft x-ray high-harmonic generation up to the nitrogen
  k-edge driven by a {kHz}, 2.1 {$\mu$m} {OPCPA} source.
\newblock {\em Journal of Physics B: Atomic, Molecular and Optical Physics},
  49(15):155601, Jul 2016.

\bibitem{Li2017}
Jie Li, Xiaoming Ren, Yanchun Yin, Kun Zhao, Andrew Chew, Yan Cheng, Eric
  Cunningham, Yang Wang, Shuyuan Hu, Yi~Wu, Michael Chini, and Zenghu Chang.
\newblock {53-attosecond X-ray pulses reach the carbon K-edge}.
\newblock {\em Nature Communications}, 8(1):186, Aug 2017.

\bibitem{Johnsoneaar3761}
Allan~S. Johnson, Dane~R. Austin, David~A. Wood, Christian Brahms, Andrew
  Gregory, Konstantin~B. Holzner, Sebastian Jarosch, Esben~W. Larsen, Susan
  Parker, Christian~S. Str\"{u}ber, Peng Ye, John W.~G. Tisch, and Jon~P.
  Marangos.
\newblock High-flux soft x-ray harmonic generation from ionization-shaped
  few-cycle laser pulses.
\newblock {\em Science Advances}, 4(5), 2018.

\bibitem{Schmidt:18}
C{\'{e}}dric Schmidt, Yoann Pertot, Tadas Balciunas, Kristina Zinchenko, Mary
  Matthews, Hans~Jakob W{\"{o}}rner, and Jean-Pierre Wolf.
\newblock {High-order harmonic source spanning up to the oxygen K-edge based on
  filamentation pulse compression}.
\newblock {\em Opt. Express}, 26(9):11834--11842, Apr 2018.

\bibitem{doi:10.1021/acs.jpclett.8b03420}
Carlo Kleine, Maria Ekimova, Gildas Goldsztejn, Sebastian Raabe, Christian
  Str\"{u}ber, Jan Ludwig, Suresh Yarlagadda, Stefan Eisebitt, Marc J.~J.
  Vrakking, Thomas Elsaesser, Erik T.~J. Nibbering, and Arnaud Rouz\'{e}e.
\newblock {Soft X-ray Absorption Spectroscopy of Aqueous Solutions Using a
  Table-Top Femtosecond Soft X-ray Source}.
\newblock {\em The Journal of Physical Chemistry Letters}, 10(1):52--58, 2019.

\bibitem{Li:19}
Jie Li, Andrew Chew, Shuyuan Hu, Jonathon White, Xiaoming Ren, Seunghwoi Han,
  Yanchun Yin, Yang Wang, Yi~Wu, and Zenghu Chang.
\newblock Double optical gating for generating high flux isolated attosecond
  pulses in the soft {X}-ray regime.
\newblock {\em Opt. Express}, 27(21):30280--30286, Oct 2019.

\bibitem{Barreau2020}
Lou Barreau, Andrew~D. Ross, Samay Garg, Peter~M. Kraus, Daniel~M. Neumark, and
  Stephen~R. Leone.
\newblock Efficient table-top dual-wavelength beamline for ultrafast transient
  absorption spectroscopy in the soft {X}-ray region.
\newblock {\em Scientific Reports}, 10(1):5773, Apr 2020.

\bibitem{gebhardt2020bright}
M.~Gebhardt, T.~Heuermann, R.~Klas, C.~Liu, A.~Kirsche, M.~Lenski, Z.~Wang,
  C.~Gaida, J.~E. Antonio-Lopez, A.~Sch\"{u}lzgen, R.~Amezcua-Correa,
  J.~Rothhardt, and J.~Limpert.
\newblock Bright, high repetition rate water window soft {X}-ray source enabled
  by nonlinear pulse self-compression in antiresonant hollow-core fibre, 2020.

\bibitem{Pertot264}
Yoann Pertot, C\'{e}dric Schmidt, Mary Matthews, Adrien Chauvet, Martin
  Huppert, Vit Svoboda, Aaron von Conta, Andres Tehlar, Denitsa Baykusheva,
  Jean-Pierre Wolf, and Hans~Jakob W\"{o}rner.
\newblock Time-resolved x-ray absorption spectroscopy with a water window
  high-harmonic source.
\newblock {\em Science}, 355(6322):264--267, 2017.

\bibitem{Attar54}
Andrew~R. Attar, Aditi Bhattacherjee, C.~D. Pemmaraju, Kirsten Schnorr,
  Kristina~D. Closser, David Prendergast, and Stephen~R. Leone.
\newblock Femtosecond x-ray spectroscopy of an electrocyclic ring-opening
  reaction.
\newblock {\em Science}, 356(6333):54--59, 2017.

\bibitem{doi:10.1021/acs.jpclett.9b03559}
Adam~D. Smith, Tadas Bal\v{c}iunas, Yi-Ping Chang, C\'{e}dric Schmidt, Kristina
  Zinchenko, Fernanda~B. Nunes, Emanuele Rossi, V^^c3^^adt Svoboda, Zhong Yin,
  Jean-Pierre Wolf, and Hans~Jakob W\"{o}rner.
\newblock Femtosecond soft-x-ray absorption spectroscopy of liquids with a
  water-window high-harmonic source.
\newblock {\em The Journal of Physical Chemistry Letters}, 11(6):1981--1988,
  2020.
\newblock PMID: 32073862.

\bibitem{PhysRevLett.98.013901}
J.~Tate, T.~Auguste, H.~G. Muller, P.~Sali\'{e}res, P.~Agostini, and L.~F.
  DiMauro.
\newblock Scaling of wave-packet dynamics in an intense midinfrared field.
\newblock {\em Phys. Rev. Lett.}, 98:013901, Jan 2007.

\bibitem{PhysRevLett.103.073902}
A.~D. Shiner, C.~Trallero-Herrero, N.~Kajumba, H.-C. Bandulet, D.~Comtois,
  F.~L\'{e}gar\'{e}, M.~Gigu\`{e}re, J-C. Kieffer, P.~B. Corkum, and D.~M.
  Villeneuve.
\newblock Wavelength scaling of high harmonic generation efficiency.
\newblock {\em Phys. Rev. Lett.}, 103:073902, Aug 2009.

\bibitem{doi:10.1063/1.5041498}
Allan~S. Johnson, David Wood, Dane~R. Austin, Christian Brahms, Andrew Gregory,
  Konstantin~B. Holzner, Sebastian Jarosch, Esben~W. Larsen, Susan Parker,
  Christian Str\"{u}ber, Peng Ye, John W.~G. Tisch, and Jon~P. Marangos.
\newblock Apparatus for soft x-ray table-top high harmonic generation.
\newblock {\em Review of Scientific Instruments}, 89(8):083110, 2018.

\bibitem{Fu2020}
Yuxi Fu, Kotaro Nishimura, Renzhi Shao, Akira Suda, Katsumi Midorikawa, Pengfei
  Lan, and Eiji~J. Takahashi.
\newblock High efficiency ultrafast water-window harmonic generation for
  single-shot soft {X}-ray spectroscopy.
\newblock {\em Communications Physics}, 3(1):92, May 2020.

\bibitem{Fu2018}
Yuxi Fu, Katsumi Midorikawa, and Eiji~J. Takahashi.
\newblock Towards a petawatt-class few-cycle infrared laser system via
  dual-chirped optical parametric amplification.
\newblock {\em Scientific Reports}, 8(1):7692, May 2018.

\bibitem{PhysRevLett.83.2187}
Charles~G. Durfee, Andy~R. Rundquist, Sterling Backus, Catherine Herne,
  Margaret~M. Murnane, and Henry~C. Kapteyn.
\newblock Phase matching of high-order harmonics in hollow waveguides.
\newblock {\em Phys. Rev. Lett.}, 83:2187--2190, Sep 1999.

\bibitem{PhysRevLett.82.1668}
E.~Constant, D.~Garzella, P.~Breger, E.~M\'{e}vel, Ch. Dorrer, C.~Le~Blanc,
  F.~Salin, and P.~Agostini.
\newblock Optimizing high harmonic generation in absorbing gases: Model and
  experiment.
\newblock {\em Phys. Rev. Lett.}, 82:1668--1671, Feb 1999.

\bibitem{7927712}
Y.~{Fu}, E.~J. {Takahashi}, and K.~{Midorikawa}.
\newblock Energy scaling of infrared femtosecond pulses by dual-chirped optical
  parametric amplification.
\newblock {\em IEEE Photonics Journal}, 9(3):1--8, 2017.

\bibitem{Xu:20}
Lu~Xu, Kotaro Nishimura, Akira Suda, Katsumi Midorikawa, Yuxi Fu, and Eiji~J.
  Takahashi.
\newblock Optimization of a multi-{TW} few-cycle 1.7-$\mathrm{\mu}$m source
  based on {Type-I BBO} dual-chirped optical parametric amplification.
\newblock {\em Opt. Express}, 28(10):15138--15147, May 2020.

\bibitem{Lai:11}
Chien-Jen Lai and Franz~X. K\"{a}rtner.
\newblock The influence of plasma defocusing in high harmonic generation.
\newblock {\em Opt. Express}, 19(23):22377--22387, Nov 2011.

\bibitem{Xu:2020}
Lu~Xu, Kotaro Nishimura, Akira Suda, Katsumi Midorikawa, Yuxi Fu, and Eiji~J.
  Takahashi.
\newblock 2.2 {TW} few-cycle intense pulses at 1.7 {$\mu$m} based on a {BBO}
  dual-chirped optical parametric amplification.
\newblock In {\em Proceedings of High-brightness Sources and Light-driven
  Interactions Congress}, 2020.
\newblock MM2C.3.

\bibitem{Sisourat2010}
Nicolas Sisourat, Nikolai~V. Kryzhevoi, P{\v{r}}emysl Koloren{\v{c}}, Simona
  Scheit, Till Jahnke, and Lorenz~S. Cederbaum.
\newblock Ultralong-range energy transfer by interatomic coulombic decay in an
  extreme quantum system.
\newblock {\em Nature Physics}, 6(7):508--511, Jul 2010.

\bibitem{Fohlisch2005}
A.~F\"{o}hlisch, P.~Feulner, F.~Hennies, A.~Fink, D.~Menzel, D.~Sanchez-Portal,
  P.~M. Echenique, and W.~Wurth.
\newblock Direct observation of electron dynamics in the attosecond domain.
\newblock {\em Nature}, 436(7049):373--376, Jul 2005.

\end{thebibliography}
\end{document}